\documentclass[a4paper]{jpconf}
\usepackage{graphicx}
\usepackage{amssymb}
\usepackage{hyperref}

\begin{document}
\title{\vspace{-2.05cm}
\hfill{\small{DESY 15-003}}\\[1.27cm]
Looking for dark matter on the light side}

\author{Babette D\"obrich}

\address{DESY, Notkestr. 85, 22607 Hamburg, Germany}

\ead{babette.doebrich@desy.de}

\begin{abstract}
Among the prominent low-mass dark matter candidates is the QCD axion
but also other light and weakly interacting particles
beyond the Standard Model.
We review briefly
the case for such dark matter and 
give an overview on most recent experimental
efforts within laboratory searches, where we focus on experiments
exploiting a potential electromagnetic coupling of such particles.
\end{abstract}

\section{Three ultra-light dark matter candidates}

It would be a huge break-through to find out what dark matter (DM) is made of.
Whilst its constituents could be rather heavy and, e.g., leave an `imprint' at
the LHC,  show up in astrophysics signatures or reveal themselves through
recoil energy at direct detection setups \cite{others},
there are well-motivated candidates also on the `light side'.
Our main concern in this article are henceforth DM candidates below the 
eV mass-scale.

Of such ultra-light particles, the QCD axion is the most
prominent dark matter candidate, see, e.g., \cite{Sikivie:1900zz} for a review.
On the {\it theoretical side}, the axion is
a pseudo-scalar pseudo-Goldstone boson that is
a consequence of the Peccei-Quinn 
solution \cite{Peccei:1977ur} to the strong CP problem. 
The strong CP
problem amounts to the question why CP violation in QCD
is unmeasurably small (or even absent). The effective parameter 
for CP violation
receives contributions from the $\theta$-angle
of QCD, being essentially unconstrained a priori,
and the quark mass matrix. 
As these parameters are unrelated from the outset, it  arises the
question for a natural explanation on why the CP-violating parameter
is so close to zero. 
In essence,
the axion solution to the strong CP-problem makes
the parameter a dynamical variable
which naturally relaxes to zero.

From an {\it experimental viewpoint}, it is most interesting that 
axions - in certain parameter regimes
- constitute a perfect candidate to make
up the cold dark matter (CDM) in our universe.
Although the axion is very light, it can be non-thermally
produced in the early universe \cite{Preskill:1982cy,Sikivie:2006ni}.

A particularly attractive feature of axion dark matter
is that its viable parameter range is comparatively small 
and it is thus a realistic aim to confirm or
exclude axions as main dark matter component with current and near-future
technology.
For axions as cold DM, two natural cosmological windows
exist, see, e.g., \cite{Hertzberg:2008wr}:
In the post-inflation scenario,
the spontaneous breaking of the Peccei-Quinn symmetry at a scale $f$, 
which gives rise to the axion 
as its pseudo Goldstone boson, takes place
only after inflation.
In the other scenario, the Peccei-Quinn phase transition
happens before inflation.
The former scenario is typically related to
axions with higher masses than the latter because the decay of axionic topological 
defects (absent in the latter) also produces DM axions and
generically the DM abundance grows with the decay constant
(i.e. decreases with the axion mass).

\begin{figure}[h]
\includegraphics[width=\textwidth]{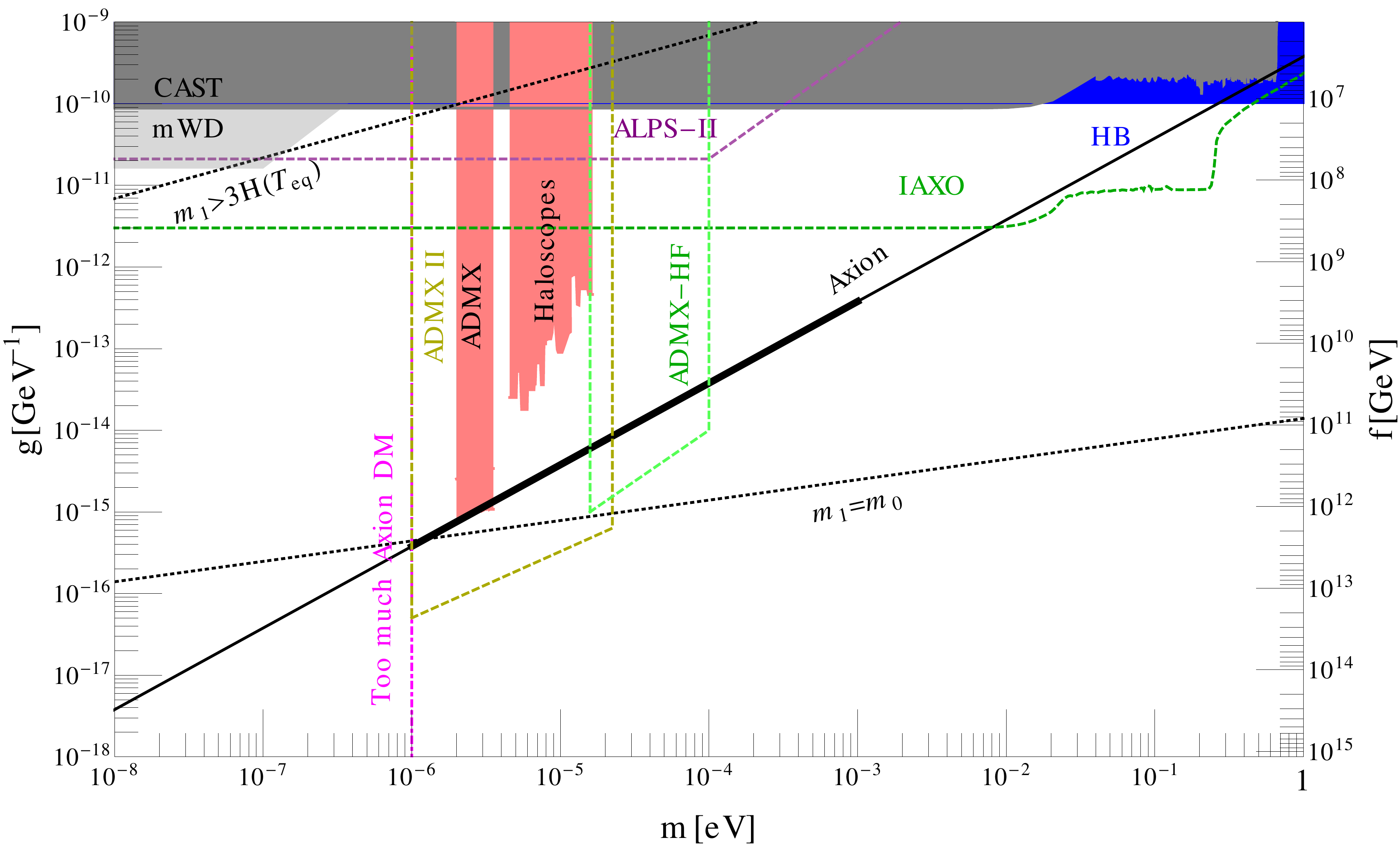}\hspace{2pc}%
\caption{\label{ALPDM}
Parameter space of axion and axion-like particle DM (photon-coupling) in the sub-eV range.
{  The QCD axion is predicted in an $\mathcal{O}(1)$
region around the black line labeled `Axion'. As Dark Matter, the axion should not be much lighter
than $10^{-6}$eV not to overproduce DM, as indicated by the magenta dot-dashed line. On the other
hand, concrete models for ALP DM have been proposed for the region round and below the dotted line
labeled $m_0=m_1$. The dotted line labeled $m_1>3H(T_{\rm eq})$ sets a model-building limit for ALP DM.
From the observational point of view, the colored regions and and dashed lines are interesting:
The colored parameter regions `mWD', `HB' and `CAST', `ADMX', `Haloscopes'
are excluded by observation and experiment, respectively.
The dashed lines indicated the reach of planned experiments and experiments under construction,
see text for details}
}
\end{figure}

 The situation for the axion is sketched in Fig.~\ref{ALPDM} where performed experiments
 and the foreseen reach of some experiments that are
 planned for current and near future are shown. In addition, some astrophysical 
 bounds are shown, for more insight on this matter see, e.g. \cite{Raffelt:2006cw,Agashe:2014kda}.
 In Fig.~\ref{ALPDM}, for the black line labeled `axion' we have implemented  $g= \frac{\alpha}{2 \pi f}$
as relation of the axion-to-photon coupling and the symmetry breaking scale.
 For specific axion models, a $\mathcal{O}(1)$ factor enters in the relation between $g$ and $f$, 
 cf.  \cite{Agashe:2014kda}. 
 Axions are a good cold dark matter candidate roughly between masses of $10^{-6}$eV and $10^{-3}$eV
\cite{Archidiacono:2013cha,Kim:1979if} (indicated by thickening the axion line in Fig.~\ref{ALPDM}.
 Below $10^{-6}$eV, axions tend to 
 produce too much DM and they are disfavored, 
 although some models can justify the selection of suitable initial conditions. 
Note that at very low masses, there might be also effects on black hole dynamics
\cite{Arvanitaki:2009fg}.
In laboratory searches, to probe 
the very tiny couplings that correspond to high values 
of the axion symmetry breaking scale,
resonant search strategies are mostly employed and 
will be briefly sketched in Sect.~\ref{sec:res}

The axion is very light and very weakly interacting. Thus it exhibits
the properties
of a more general class of particles dubbed {\it `Weakly Interacting Slim 
Particles' (WISPs)} \cite{Jaeckel:2010ni,Baker:2013zta,Essig:2013lka}.
Among the WISPs that could be cold dark matter \cite{Arias:2012az} are also
axion-like particles (ALPs). Such general ultralight pseudo-scalars
could be pseudo-Nambu Goldstone bosons (pNGB) associated with
a symmetry breaking scale different from the Peccei-Quinn scale \cite{Ringwald:2012hr,Ringwald:2013via}. Effectively they 
can have a coupling to photons similarly to the QCD axion, 
but a `relaxed' mass-coupling relation (i.e.,
they are not confined to the parameter region around the
axion line in Fig.~\ref{ALPDM}). 
Their existence is motivated in Standard Model extensions \cite{Jaeckel:2010ni},
and they have been also evoked to explain some
astrophysical puzzles such as the observed transparency of the universe to high-energetic
photons, see, e.g., \cite{Horns:2012fx}.

Similar to axions, ALP CDM \cite{Arias:2012az} can be realized through a {\it misalignment
mechanism}: The axion mass $m$ is strongly temperature dependent (same might apply to ALPs as pNGBs).
For axions above the symmetry breaking scale (in the early universe), 
the axion is essentially massless. Thus at very high
energies, the initial angle is not fixed (for the axion it needs 
not to be at its CP conserving value), i.e., it can be mis-aligned from
its minimum. The equation of motion
for the field in the expanding universe is that of a damped harmonic oscillator (where $3 H$ quantifies
the damping term and $H$ is the Hubble expansion parameter) and initially
the field is frozen when $3H \gg m$.
At later times $t_1$, when the particle mass $m_1= 3 { H}$, the  field
starts to oscillate and behaves as cold dark matter fluid.
The situation for ALP DM is reviewed in part through the dotted black lines
in Fig.~\ref{ALPDM}:
The upper line, labeled $m_1> 3 { H} (T_{\rm eq})$ is an upper bound for any such
ALP DM model: The mass $m_1$ at the which the  oscillations
start should be attained latest at matter-radiation equality. 
The lower black dotted line labeled $m_1=m_0$ denotes `Standard ALP DM' and
is the simplest ALP DM
model in which $m_1$ is the same as the
ALP mass today ($m_0$). Models in which $m_1 \ll m_0$ can in
principle create a sufficient DM abundance at slightly higher coupling values \cite{Arias:2012az}, but
most model-building efforts favor ALP DM within roughly an order of magnitude above this line. 

Beyond ALPs, {\it hidden photons} (HPs, reviewed, e.g., in \cite{Jaeckel:2013ija}),
which are hidden sector U(1)s coupled kinetically to
the photons of the visible sector, are WISPy cold
dark matter candidates \cite{Nelson:2011sf,Arias:2012az}.
Such particles would - with a hidden Higgs or 
St\"uckelberg generated mass term -
be manifest in photon hidden-photon
oscillations similar to what is observed with neutrinos.
Fig.~\ref{HPDM} shows the viable parameter space for HP DM
according to \cite{Arias:2012az}, in which the orange region labeled `Xenon' 
denotes limits inferred
from the XENON10 experiment \cite{An:2013yua} (see also \cite{Vinyoles:2015aba} for novel bounds
on the longitudinal HP component).

In summary: For both ALPs and hidden photons the parameter regime in which they 
can constitute DM
is much larger than for QCD axions.
Thus, in laboratory searches, also 
non-resonant techniques have become attractive as they allow faster scanning
(albeit at reduced overall sensitivity). 
This is discussed
in Sect~\ref{sec:nonres}.
In the following we give a brief overview on laboratory searches
for axion and WISP DM. 
We focus on searches exploiting coupling to light.
Recent reviews on ultra-light particle dark matter
with a more theory-based focus include \cite{Kawasaki:2013ae,Ringwald:2013via}.

\section{Laboratory setups looking for ultra-light dark matter}

It might be helpful to first
categorize experiments that are sensitive to light \& weakly interacting 
particles of the above described types in the following manner:

Firstly, there are setups which could find such particles outside the range 
in which they likely constitute dark matter in the most immediate models
{ and will thus not concern us in the following.}
This is because these setups cannot probe small enough couplings.
{ The experimental limits of these setups are well below the
sensitivity of the filled (and thereby excluded) regions
in Fig.~\ref{ALPDM}, but can be argued to be less model-dependent (see, e.g.
\cite{Baker:2013zta,Essig:2013lka} for a comprehensive review of the WISP parameter spaces at
larger coupling values).}

Secondly, there are experiments which could find these 
particles inside the range in which they
could be cold dark matter, however independent of 
the fact whether they actually are dark matter (indirect searches).
An example is second-generation light-shining-through-a-wall LSW \cite{Redondo:2010dp}
like ALPS-II \cite{Bahre:2013ywa}.
Another example are Helioscopes, that aim to 
convert axions or the like emitted from our
sun into a detectable X-ray signal,
such as with the Cern Axion Solar
Telescope (CAST) (\cite{Arik:2013nya} for recent results, grey in Fig.~\ref{ALPDM}) or the future
International Axion Observatory (IAXO) \cite{Armengaud:2014gea}.
In Fig.~\ref{ALPDM} the foreseen reach of ALPS-II and IAXO is shown in dashed pink and green
in the upper left,
respectively.
{ The grayish region labeled `mWD' shows an astrophysical exclusion
according to \cite{Gill:2011yp}.}

Thirdly, there are experiments which are conceived such that they (probably) 
find only  something
if the above described particles are dark matter.
In the following, we mention only such recent and planned setups/ideas.
As motivated above, we group the searches into
`resonant' and `non-resonant' although
some setups can do both sort of scans.
The former technique has an advantage
of highest sensitivity, the latter of an increased scanning speed.

\subsection{Resonant searches \label{sec:res}}

A resonant Haloscope is the most prominent and thus far most
sensitive experiment searching for ultra-light dark matter.
Based on the concept suggested in \cite{Sikivie:1983ip}, 
several such experiments have been performed, see
\cite{Rybka:2014xca,Stern:2014wma} for recent reviews:
A narrow band microwave resonator is placed in a (typically solenoidal)
strong magnetic field. If axions constitute dark matter, they should then
stream in large numbers through this experimental apparatus.
In the magnetic field, axions would then be converted into photons
and be visible at the resonant frequency of the cavity (related to the 
DM particle mass).
Here, the conversion probability of the axions 
into photons is increased by the power built-up factor of the resonator. 
Hence, the sensitivity of such experiments can be 
increased by several orders of magnitudes compared to non-resonant techniques.
Due to the pseudo-scalar nature of the axion, the electric fields
of the cavity-mode and the external magnetic field should have a sizable overlap.
In Fig.~\ref{ALPDM}, obtained Haloscope
limits are shown in orange (labeled `Haloscopes'
and `ADMX') \cite{Hagmann:1990tj,Wuensch:1989sa,DePanfilis:1987dk,Asztalos:2003px,Asztalos:2001jk,Asztalos:2001tf,Asztalos:2009yp}.
{ ADMX has searched for  KSVZ axions, 
over the 1.9 - 3.6 $\mu$eV mass range, whereas other performed 
haloscope searches were so far strictly only sensitive to ALP DM.}
Currently
two further, more sensitive Haloscope measurements are prepared at Seattle (ADMX II) and
Yale (ADMX-HF). Prospects are adapted from \cite{Shokair:2014rna}
in the yellow and green dashed areas of Fig.~\ref{ALPDM}, respectively.
{ Note that these prospects and the limits are shown under the assumption $\Omega_a=\Omega_{\rm DM}$,
i.e. that axions make up all of the Dark Matter.}

For Haloscope searches in the solenoid configuration, 
it eventually can become difficult at the high mass range
to keep the cavity resonant whilst retaining
a sizable cavity volume (high mass $\rightarrow$ smaller length scales)
\footnote{\& at the same time
detection techniques need careful study at higher masses  \cite{Lamoreaux:2013koa}.}.
Thus, for {\it higher-mass DM } axion searches, additional
techniques have been suggested and corresponding studies are underway.
An example is the ORPHEUS setup \cite{Rybka:2014cya}
based on wire-planes in a confocal resonator. 
The wire-planes allow to alternate the B-field
direction and keep a sizable overlap of external magnetic field
and electric photon fields and thus probe higher frequencies  \cite{Sikivie:1993jm}.
Besides these efforts, let us also point to the emerging activities at
the `Center for Axion and Precision Physics Research' (CAPP) in Korea \cite{capp}
from where novel contributions to DM axion detection are emerging currently.

A possibility for either resonant (or broadband, see below) direct dark matter searches
might open also up with 
IAXO \cite{Armengaud:2014gea}.
IAXO, although its primary mission is to act as a Helioscope like described above,
could host, e.g., a cavity set-up in one of its bores. IAXO
will be based on a toroidal magnet. Thus, cylindrical cavities
do not suggest themselves. However, e.g., a rectangular cavity structure \cite{Baker:2011na,Irastorza:2012jq}
is conceivable. { In principle,
very large cavities could be feasible to explore a mass
range below ADMX (making use of the
huge magnetic volume of IAXO).  This is however a limited possibility
amongst else because it would likely interfere with the original purpose of IAXO as a helioscope.}

If long thin cavities
are realized, this opens up possibilities also for the higher mass-range  \cite{javig}.
For testing such rectangular cavities before IAXO exists, e.g., the CAST magnet 
as well as a straightened\footnote{For HERA magnets, the cold-mass (including the 
beam pipe) was originally fabricated straight and then bent by its outer shell.
It has been shown by the ALPS-II collaboration
that the beam-pipe can be straightened to
almost the full aperture \cite{Bahre:2013ywa} and thus 
fit appropriate test cavities (the CAST magnet has a straight beam
pipe to start with). { In its final stage, ALPS-II 
plans to use 20 HERA magnets, whose later use in DM search seems to be an intriguing possibility also}.}
 HERA dipole magnet could be an appropriate test-bench.
{ In fact R\&D in this direction has started on a small scale in Valencia  \cite{javig}.
A 1m long $1\times2$cm cavity (resonant at about 60 $\mu$eV) is being characterized towards this purpose.
For a single HERA magnet, the magnetic length is 8.83m at 5.3T,
for CAST, the magnetic length  is 14.3m with 9T field. Albeit the sensitivity
reach of these test setups depends on results
for the cavity quality factor the noise detection level,
such relatively quick R\&D measurements will be crucial in determining the real DM
potential of IAXO.}

Another Haloscope experiment focusing (so far) on 
HP DM also exists at DESY, called WISPDMX \cite{Horns:2013ira}. It employs a former HERA
proton cavity and is most sensitive in the 200-600 MHz range and
can also run in `broadband mode', see \cite{Horns:2014qta}.
The `broadband mode' is possible
since for HPs no mode-overlap with an external field is needed.

A recent suggestion that could complement cavity searches at {\it very low masses}
is the resonant axion DM detection through cryogenic LC-circuits in
a strong external magnetic field \cite{Sikivie:2013laa}:
axion CDM could induce a current in the part of an LC circuit that is immersed in an external
magnetic field.
Nicely, such setups are also apt to scan a large part of the HP cold DM parameter space, see
\cite{Arias:2014ela,Chaudhuri:2014dla}, although a careful design is likely needed
that accounts for the proper shielding of the setup:
As HP DM does not need the magnetic field for its conversion in a photon
signal, the HP DM stream will also excite electrons in the shield,
causing a parasitic contribution that needs to be accounted for in the experiment's 
design \cite{Chaudhuri:2014dla}.

So far we have have described setups that exploit coupling of the
light dark matter particles to photons only. Of course
there are alternative techniques, focusing on the coupling of axions to other particles.
Examples of recent works point out that axion DM could be found through time-varying
CP-odd nuclear moments \cite{Graham:2013gfa,Budker:2013hfa}, see also \cite{Roberts:2014dda},
atomic transitions \cite{Sikivie:2014lha} or
NMR techniques \cite{Arvanitaki:2014dfa}.
As mentioned, we focus on
DM-to-photon couplings in the following and turn to the non-resonant techniques now.

\begin{figure}[h]
\includegraphics[width=20pc]{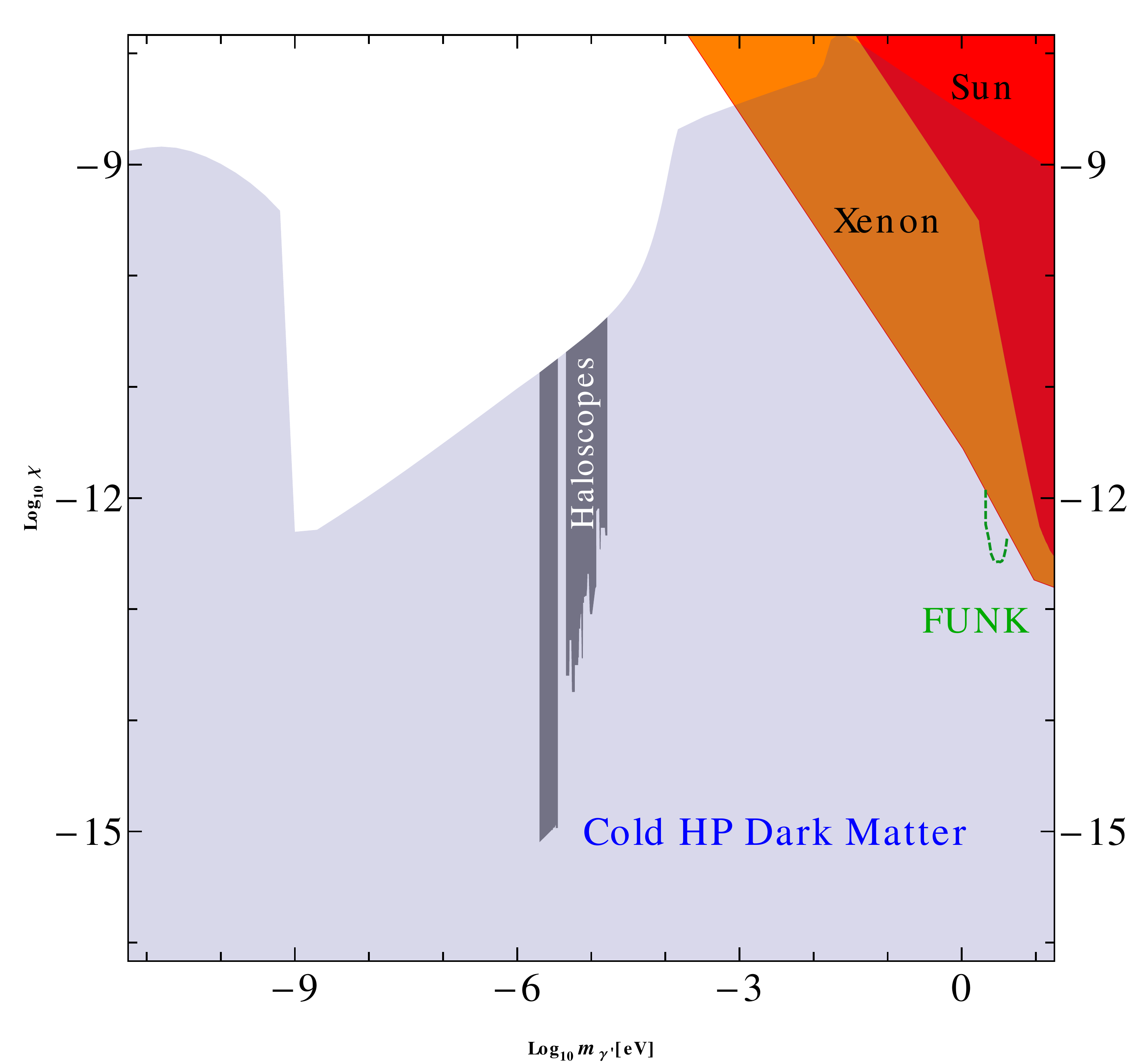}\hspace{2pc}%
\begin{minipage}[b]{14pc}\caption{\label{HPDM}
{ Parameter space (coupling ${\rm Log}_{10} \ \chi$ in vs mass ${\rm Log}_{10} \ m_{\gamma'}$ [eV])
of hidden photon dark matter in the sub-eV range.
The blue shaded region { labeled `Cold HP Dark Matter'} is the viable HP DM parameter space as discussed in  \cite{Arias:2012az}.
The dashed green region shows the parameter reach envisaged by the non-resonant FUNK experiment
in Karlsruhe in a first search in the optical (searches at other frequencies are conceivable, see \cite{Dobrich:2014kda}).}
Other regions see text.}
\end{minipage}
\end{figure}

\subsection{Broadband search \label{sec:nonres}}

As has become obvious from the above considerations, the resonant enhancement
is a crucial factor in becoming sensitive to very small couplings and thus eventually
 to QCD axions. On the other hand, 
 it can be favorable to abstain from tuning to resonance, thereby losing
sensitivity in coupling, but gaining in the width of the mass-range 
that can be covered
within a fixed amount of 
time\footnote{This assumes that appropriate low-noise
broad-band detectors are available
which
can be difficult  over the entire spectral range of interest.}.

Such a concept was proposed in \cite{Horns:2012jf,Jaeckel:2013eha} as DM search with a dish antenna:
axions and ALP DM (in combination with a magnetic field) as well as HP DM can excite electrons
at the surface of a conductor, owing to their (albeit very weak) electromagnetic coupling to them.
If this conductor is a spherical mirror, a consequence of this process would
a photonic signal at the center of the mirror sphere (whereas most background
photons are scattered to the focal point).
Such a setup has a directional sensitivity \cite{Jaeckel:2013sqa}
as the exact position of a signal spot will depend on the 
velocity distribution
of the DM
with respect to the dish.
A global off-set of the signal 
can be expected due to the movement of the sun in the galactic rest frame
as well as a daily modulation on the same order of magnitude
(the yearly modulation is negligible due to the 
small velocity of the earth around the sun).
Besides the signal-spot movement,
a velocity distribution
of the DM leads to a broadening of the signal spot.
Such considerations can ultimately 
help to verify the dark matter nature of a signal.
A dish dark matter search (Finding U(1)'s of a Novel Kind - FUNK) 
is currently being set up \cite{Dobrich:2014kda}
at Karlsruhe. It uses a prototype mirror
built for the fluorescence detector for the Pierre Auger observatory. Fig.~\ref{HPDM}
shows a plausible parameter reach for a first measurement in the optical in green.

{ Searching axion and ALP DM with the Dish technique is much more involved.
For such an experiment, the mirror would have to be magnetized {\it along} the surface (for a setting
 $\vec{B} \parallel \vec{E}$ the conversion to pseudoscalars is maximized). 
For example, a magnet with reasonable field strengths $\mathcal{O}(1)$ Tesla
that can host the Karlsruhe mirror would be difficult to procure and the experiment would become considerably
more complicated.
The parameter reach for ALP DM that could be expected in principle with the Dish technique is shown in \cite{Horns:2012jf},
however, as we are not aware of detailed proposals for such setups,
we have not included this in Fig.~\ref{ALPDM}.}

\section{Summary}

In this talk, we have briefly reviewed the case for dark matter
candidates below the eV-scale, notably axions, axion-like-particles and hidden photons.
We have reviewed recent experiments thriving to find them through their effective coupling
to electromagnetic waves. As of today, dark matter could be light, medium mass
or heavy,
a single species or a mixture of particles. Thus
it seems well advised to have a well-balanced variety of experimental searches
until we learn more about it. For dark matter candidates in
the sub-eV regime, in view of the upcoming experiments
described above, we are entering very interesting times.

\ack
I would like to thank the organizers of the 24th European Cosmic Ray 
Symposium in Kiel
for the kind invitation to present the case
for light dark matter and the chance
to learn about a variety of topics beyond my field.

 I cordially thank A.~Lindner, J.~Redondo and A.~Ringwald
 for 
 fruitful discussion and 
collaboration on the above topics and 
helpful feed-back on this note. 
In addition, I would like to thank K.~v.~Bibber for providing some of the sensitivity data.


\section*{References}
\begin{thebibliography}{9}

\bibitem{others}
See the contributions of Lars Bergstrom, Roberto Battiston and Gernot 
Maier to these proceedings

\bibitem{Sikivie:1900zz} 
  P.~Sikivie,
  ``Axions'',
  In `Particle dark matter: 
  observations, models and searches',  Bertone, G. (ed.),  Cambridge University Press, 2010 pp. 204-227
  
  
\bibitem{Peccei:1977ur} 
  R.~D.~Peccei and H.~R.~Quinn,
  Phys.\ Rev.\ D {\bf 16}, 1791 (1977).
  R.~D.~Peccei and H.~R.~Quinn,
  Phys.\ Rev.\ Lett.\  {\bf 38}, 1440 (1977).
  
    
  \bibitem{Preskill:1982cy}
  J.~Preskill, M.~B.~Wise and F.~Wilczek,
  Phys.\ Lett.\ B {\bf 120} (1983) 127;   L.~F.~Abbott and P.~Sikivie,
  Phys.\ Lett.\ B {\bf 120} (1983) 133;
  M.~Dine and W.~Fischler,
  Phys.\ Lett.\ B {\bf 120} (1983) 137.
  

  
\bibitem{Sikivie:2006ni} 
  P.~Sikivie,
  Lect.\ Notes Phys.\  {\bf 741}, 19 (2008)
  [astro-ph/0610440].
  
\bibitem{Hertzberg:2008wr} 
  M.~P. Hertzberg, M.~Tegmark and F.~Wilczek,
  Phys.\ Rev.\ D {\bf 78}, 083507 (2008)
  [arXiv:0807.1726 [astro-ph]].

  

  
      
\bibitem{Raffelt:2006cw} 
  G.~G.~Raffelt,
  Lect.\ Notes Phys.\  {\bf 741}, 51 (2008)
  [hep-ph/0611350].

  




\bibitem{Agashe:2014kda} 
  K.~A.~Olive {\it et al.}  [Particle Data Group Collaboration],
  Chin.\ Phys.\ C {\bf 38}, 090001 (2014).
  
\bibitem{Archidiacono:2013cha} 
  M.~Archidiacono, S.~Hannestad, A.~Mirizzi, G.~Raffelt and Y.~Y.~Y.~Wong,
  JCAP {\bf 1310}, 020 (2013)
  [arXiv:1307.0615 [astro-ph.CO]].




\bibitem{Kim:1979if} 
  J.~E.~Kim,
  Phys.\ Rev.\ Lett.\  {\bf 43}, 103 (1979);
  
  M.~A.~Shifman, A.~I.~Vainshtein and V.~I.~Zakharov,
  Nucl.\ Phys.\ B {\bf 166}, 493 (1980).
  
\bibitem{Arvanitaki:2009fg} 
  A.~Arvanitaki, S.~Dimopoulos, S.~Dubovsky, N.~Kaloper and J.~March-Russell,
  Phys.\ Rev.\ D {\bf 81}, 123530 (2010)
  [arXiv:0905.4720 [hep-th]].
  

\bibitem{Jaeckel:2010ni} 
  J.~Jaeckel and A.~Ringwald,
  Ann.\ Rev.\ Nucl.\ Part.\ Sci.\  {\bf 60}, 405 (2010)
  [arXiv:1002.0329 [hep-ph]].


\bibitem{Baker:2013zta} 
  K.~Baker  {\it et al.}
  Ann. Phys. (Berlin) 525, No. 6, A93-A99 (2013)
  [arXiv:1306.2841 [hep-ph]].
  
    
\bibitem{Essig:2013lka} 
  R.~Essig, 
  {\it et al.},
  arXiv:1311.0029 [hep-ph].
  



\bibitem{Arias:2012az} 
  P.~Arias, D.~Cadamuro, M.~Goodsell, J.~Jaeckel, J.~Redondo and A.~Ringwald,
  JCAP {\bf 1206}, 013 (2012)
  [arXiv:1201.5902 [hep-ph]].




  
\bibitem{Ringwald:2012hr} 
  A.~Ringwald,
  Phys.\ Dark Univ.\  {\bf 1}, 116 (2012)
  [arXiv:1210.5081 [hep-ph]].
  
\bibitem{Ringwald:2013via} 
  A.~Ringwald,
  arXiv:1310.1256 [hep-ph].
  
  
\bibitem{Horns:2012fx} 
  D.~Horns and M.~Meyer,
  JCAP {\bf 1202}, 033 (2012)
  [arXiv:1201.4711 [astro-ph.CO]].
  
\bibitem{Jaeckel:2013ija} 
  J.~Jaeckel,
  Frascati Phys.\ Ser.\  {\bf 56}, 172 (2012)
  [arXiv:1303.1821 [hep-ph]].
  
    
\bibitem{Nelson:2011sf} 
  A.~E.~Nelson and J.~Scholtz,
  Phys.\ Rev.\ D {\bf 84}, 103501 (2011)
  [arXiv:1105.2812 [hep-ph]].
  
    
\bibitem{An:2013yua} 
  H.~An, M.~Pospelov and J.~Pradler,
  Phys.\ Rev.\ Lett.\  {\bf 111}, 041302 (2013)
  [arXiv:1304.3461 [hep-ph]].
  
  
\bibitem{Vinyoles:2015aba} 
  N.~Vinyoles, A.~Serenelli, F.~L.~Villante, S.~Basu, J.~Redondo and J.~Isern,
  arXiv:1501.01639 [astro-ph.SR].

  
  
  
\bibitem{Kawasaki:2013ae} 
  M.~Kawasaki and K.~Nakayama,
  Ann.\ Rev.\ Nucl.\ Part.\ Sci.\  {\bf 63}, 69 (2013)
  [arXiv:1301.1123 [hep-ph]].


  
  
\bibitem{Redondo:2010dp} 
  J.~Redondo and A.~Ringwald,
  Contemp.\ Phys.\  {\bf 52}, 211 (2011)
  [arXiv:1011.3741 [hep-ph]].
  
  
  
  





\bibitem{Bahre:2013ywa} 
  R.~B\"ahre,   {\it et al.}
  JINST {\bf 8}, T09001 (2013)
  [arXiv:1302.5647 [physics.ins-det]].


  
  

\bibitem{Arik:2013nya} 
  M.~Arik {\it et al.}  [CAST Collaboration],
  Phys.\ Rev.\ Lett.\  {\bf 112}, no. 9, 091302 (2014)
  [arXiv:1307.1985 [hep-ex]].
  S.~Aune {\it et al.}  [CAST Collaboration],
  Phys.\ Rev.\ Lett.\  {\bf 107}, 261302 (2011)
  [arXiv:1106.3919 [hep-ex]].




\bibitem{Armengaud:2014gea} 
  E.~Armengaud {\it et al.},
  JINST {\bf 9}, T05002 (2014)
  [arXiv:1401.3233 [physics.ins-det]].
  

\bibitem{Gill:2011yp} 
  R.~Gill and J.~S.~Heyl,
  Phys.\ Rev.\ D {\bf 84}, 085001 (2011)
  [arXiv:1105.2083 [astro-ph.HE]].

  
  
  
\bibitem{Sikivie:1983ip} 
  P.~Sikivie
  Phys.\ Rev.\ Lett.\  {\bf 51}, 1415 (1983)
  [Erratum-ibid.\  {\bf 52}, 695 (1984)].
  
\bibitem{Rybka:2014xca} 
  G.~Rybka,
  Phys.\ Dark Univ.\  {\bf 4}, 14 (2014).
  



\bibitem{Stern:2014wma} 
  I.~P.~Stern,
  AIP Conf.\ Proc.\  {\bf 1604}, 456 (2014)
  [arXiv:1403.5332 [physics.ins-det]].
  
\bibitem{Hagmann:1990tj} 
  C.~Hagmann, P.~Sikivie, N.~S.~Sullivan and D.~B.~Tanner,
  Phys.\ Rev.\ D {\bf 42}, 1297 (1990).
  
\bibitem{Wuensch:1989sa} 
  W.~Wuensch, S.~De Panfilis-Wuensch, Y.~K.~Semertzidis, J.~T.~Rogers, A.~C.~Melissinos, H.~J.~Halama, B.~E.~Moskowitz and A.~G.~Prodell {\it et al.},
  Phys.\ Rev.\ D {\bf 40}, 3153 (1989).

  
  
\bibitem{DePanfilis:1987dk} 
  S.~De Panfilis, A.~C.~Melissinos, B.~E.~Moskowitz, J.~T.~Rogers, Y.~K.~Semertzidis, W.~Wuensch, H.~J.~Halama and A.~G.~Prodell {\it et al.},
  Phys.\ Rev.\ Lett.\  {\bf 59}, 839 (1987).
  
  
    
\bibitem{Asztalos:2003px} 
  S.~J.~Asztalos {\it et al.}  [ADMX Collaboration],
  Phys.\ Rev.\ D {\bf 69}, 011101 (2004)
  [astro-ph/0310042].
    
  

\bibitem{Asztalos:2001jk} 
  S.~J.~Asztalos {\it et al.}  [ADMX Collaboration],
  Astrophys.\ J.\  {\bf 571}, L27 (2002)
  [astro-ph/0104200].
  
  


  
\bibitem{Asztalos:2001tf} 
  S.~J.~Asztalos {\it et al.}  [ADMX Collaboration],
  Phys.\ Rev.\ D {\bf 64}, 092003 (2001).


  
  
\bibitem{Asztalos:2009yp} 
  S.~J.~Asztalos {\it et al.}  [ADMX Collaboration],
  Phys.\ Rev.\ Lett.\  {\bf 104}, 041301 (2010)
  [arXiv:0910.5914 [astro-ph.CO]].


  
  

  
  
  
  
  
  




  


  
  
\bibitem{Shokair:2014rna} 
  T.~M.~Shokair, J.~Root, K.~A.~Van Bibber, B.~Brubaker, Y.~V.~Gurevich, S.~B.~Cahn, S.~K.~Lamoreaux and M.~A.~Anil {\it et al.},
  Int.\ J.\ Mod.\ Phys.\ A {\bf 29}, 1443004 (2014)
  [arXiv:1405.3685 [physics.ins-det]].
  
    
\bibitem{Lamoreaux:2013koa} 
  S.~K.~Lamoreaux, K.~A.~van Bibber, K.~W.~Lehnert and G.~Carosi,
  Phys.\ Rev.\ D {\bf 88}, no. 3, 035020 (2013)
  [arXiv:1306.3591 [physics.ins-det]].
  


  
\bibitem{Rybka:2014cya} 
  G.~Rybka and A.~Wagner,
  arXiv:1403.3121 [physics.ins-det].
  
\bibitem{Sikivie:1993jm} 
  P.~Sikivie, D.~B.~Tanner and Y.~Wang,
  Phys.\ Rev.\ D {\bf 50}, 4744 (1994)
  [hep-ph/9305264].
  
  
\bibitem{capp}
\url{http://capp.ibs.re.kr/html/capp_en/}

  
  

\bibitem{Baker:2011na} 
  O.~K.~Baker  {\it et al.}
  Phys.\ Rev.\ D {\bf 85}, 035018 (2012)
  [arXiv:1110.2180 [physics.ins-det]];
  M.~Betz, F.~Caspers and K.~Zioutas,  ATS/Note/2011/066 (2011).
  
\bibitem{Irastorza:2012jq} 
  I.~G.~Irastorza and J.~A.~Garcia,
  JCAP {\bf 1210}, 022 (2012)
  [arXiv:1207.6129 [physics.ins-det]].
  
  
   \bibitem{javig}
J.~Redondo, I.~G.~Irastorza, B.~Gimeno, Expression of Interest
to the CAST steeing committee, 2014.
See also: presentation by J.~Redondo at the `IBS-MultiDark Joint focus program' in Korea, 2014
\url{http://wwwth.mpp.mpg.de/members/redondo/data/seminarslides/20141023.pdf}

  



  
\bibitem{Horns:2013ira} 
  D.~Horns, A.~Lindner, A.~Lobanov and A.~Ringwald,
  arXiv:1309.4170 [physics.ins-det].

  
\bibitem{Horns:2014qta} 
  D.~Horns, A.~Lindner, A.~Lobanov and A.~Ringwald,
  arXiv:1410.6302 [hep-ex].

  
  


%

  
\bibitem{Sikivie:2013laa} 
  P.~Sikivie, N.~Sullivan and D.~B.~Tanner,
  Phys.\ Rev.\ Lett.\  {\bf 112}, 131301 (2014)
  [arXiv:1310.8545 [hep-ph]].


\bibitem{Arias:2014ela} 
  P.~Arias, A.~Arza, B.~D\"obrich, J.~Gamboa and F.~Mendez,
  arXiv:1411.4986 [hep-ph].


\bibitem{Chaudhuri:2014dla} 
  S.~Chaudhuri, P.~W.~Graham, K.~Irwin, J.~Mardon, S.~Rajendran and Y.~Zhao,
  arXiv:1411.7382 [hep-ph].
  
  





\bibitem{Graham:2013gfa} 
  P.~W.~Graham and S.~Rajendran,
  Phys.\ Rev.\ D {\bf 88}, 035023 (2013)
  [arXiv:1306.6088 [hep-ph]].
  
\bibitem{Budker:2013hfa} 
  D.~Budker {\it et al.}
  Phys.\ Rev.\ X {\bf 4}, 021030 (2014)
  [arXiv:1306.6089 [hep-ph]].
  
    
\bibitem{Roberts:2014dda} 
  B.~M.~Roberts, Y.~V.~Stadnik, V.~A.~Dzuba, V.~V.~Flambaum, 
  N.~Leefer and D.~Budker,
  Phys.\ Rev.\ Lett.\  {\bf 113}, 081601 (2014)
  [arXiv:1404.2723 [hep-ph]].
  
    
  
\bibitem{Sikivie:2014lha} 
  P.~Sikivie,
  Phys.\ Rev.\ Lett.\  {\bf 113}, no. 20, 201301 (2014)
  [arXiv:1409.2806 [hep-ph]].

  
  \bibitem{Arvanitaki:2014dfa} 
  A.~Arvanitaki and A.~A.~Geraci,
  Phys.\ Rev.\ Lett.\  {\bf 113}, 161801 (2014)
  [arXiv:1403.1290 [hep-ph]].



  

 



\bibitem{Horns:2012jf} 
  D.~Horns  {\it et al.},
  JCAP {\bf 1304}, 016 (2013)
  [arXiv:1212.2970].

  
\bibitem{Jaeckel:2013eha} 
  J.~Jaeckel and J.~Redondo,
  Phys.\ Rev.\ D {\bf 88}, 115002 (2013)
  [arXiv:1308.1103 [hep-ph]].

\bibitem{Jaeckel:2013sqa} 
  J.~Jaeckel and J.~Redondo,
  JCAP {\bf 1311}, 016 (2013)
  [arXiv:1307.7181].
  

%
%

%
%


  


\bibitem{Dobrich:2014kda} 
  B.~D\"obrich, K.~Daumiller, R.~Engel  {\it et al.},
  arXiv:1410.0200 [physics.ins-det];
  J.~Redondo and  B.~D\"obrich,
  arXiv:1311.5341 [hep-ph].









\end{thebibliography}
\end{document}